\documentclass[12pt]{article}

\usepackage[top=2cm,bottom=2cm,left=2cm,right=2cm]{geometry}
\usepackage{bm}
\usepackage{amsmath}
\usepackage{amssymb}
\usepackage{graphicx}
\usepackage{multirow}
\usepackage{subfigure}

\title{
Universal bridge functional for infinitely diluted solutions: a case
study for Lennard-Jones spheres of different diameter 
{\small Preprint. Original (in Russian) was sent to the  Russian Journal of Physical Chemistry A. }
}
\author{Volodymyr P. Sergiievskyi, \\Andrey I. Frolov \\ Max Planck Institute for Mathematics in the Sciences,\\ Inselstrasse 22, DE-04103, Leipzig, Germany,\\ e-mail: sergiiev@mis.mpg.de}

\newcommand{\EquationRef}[2]{#1 \eqref{#2}}
\newcommand{\FigureRef}[2]{#1 \ref{#2}}
\newcommand{\TableRef}[2]{#1 \ref{#2}}

\begin{document}


\maketitle

\begin{abstract}

In the paper we propose an universal bridge functional for the closure of the Ornstein-Zernike (OZ) equation for the case of infinitely diluted solutions of Lennard-Jones shperes of different size in the Lennard-Jones fluid.
Bridge functional is paprameterized using the data of the Molecular Dynamics (MD) simulations. 
We show that for all investigated systems the bridge functional can be efficiently papameterized with the exponential function which depends only on the ratio of sizes of the solute and solvent atoms.
To check the parameterization we solve the OZ equation with the closure which includes the parametrized functional and with the closure without the bridge functional (Hyper-netted chain closure).
We show that introducing the bridge functional allows to obtain radial distribution functions (RDFs), which are close to the MD results and essentially improve predictions of the location and height of the first peak of the RDF.

\vspace{5pt}


\end{abstract}

\section{Introduction}

Integral Equation Theory Of Liquids (IETL) is an effective instrument for desciption of structural and thermodynamic properties of solutions.
The main equation of IETL is the Ornstein-Zernike integral equation \cite{Ornstein1914}, which connects the direct and the total correlation functions of the particels of the system.
Another fundamental equation which connects the direct and the total correlation functions is the closure relation \cite{Hirata2003}.
Ornstein-Zernike equation togather with the closure relation allows to calculate the particle-to-particle correlation functions of the system which in turn gives an opportunity to calculate the main thermodynamic parameters of the system \cite{Hansen2000}.
For the molecular systems one often uses approximations of the Ornstein-Zernike equation. 
The most popular of them are the reference interaction site model (RISM) \cite{Chandler1972,Hirata2003}  and three-dimensional reference interaction site model (3DRISM) \cite{Beglov1995,Kovalenko2003}. 
There were developed efficient numerical algorithms for solving the RISM and 3DRISM equations \cite{Sergiievskyi2011, Fedorov2008b, Chuev2004, Kovalenko1999b}. Recently there were proposed number of methods of parameterizing the results of RISM and 3DRISM calculations. These methods allow to predict the free energy of solvation of organic and bioactive molecules with the average error 1 kcal/mol \cite{Ratkova2010, Palmer2010a, Palmer2011,Ratkova2011,Frolov2011,Sergiievskyi2011a}.

Ornstein-Zernike, RISM and 3DRISM equations give a qualitativaly correct description of the local stucture of the liquid. For example, RISM equations can correctly predict assymetry of ion solvation \cite{Fedorov2007,Kolombet2010a}, allow to predict the stability of molecular agregates in solution \cite{Chuev2009,Yamazaki2010}, and ligand-substrate binding \cite{Genheden2010,Kolombet2010}.
However, often OZ, RISM and 3DRISM equations cannot reproduce many quantitative parameters such as the radial distribution functions (RDF) or the solvation free energy (SFE) \cite{Ten-no2010,Ratkova2011,Karino2010}.

This is explaied by the fact, that the closure relation contains so-called bridge functional which is represented as an infinite sum of integrals of correlation functions and thus is practically incomputable \cite{Hansen2000}.
Usually, to obtain the numerical solution of the equations of IETL one uses the empirical closures. Not all of these closures give the good coincidence with experiments.
Several closures was obtained for the simplified models.
To name a few: the Hypper Netted Chain (HNC) closure, where the bridge functional is simply ignored, the Percus-Yevick closure \cite{Percus1958}, the Martynov-Sarkisov closure \cite{Martynov1983}, the Verlet Modified closure \cite{Labik1991} and others \cite{Martynov1992}. 
Parameterization of the molecular dynamics (MD) simulation data is another perspective method to obtaine the bridge functional \cite{Du2000,Francova2011}.
In the paper \cite{Kovalenko2000b} it was proposed to use the closure  with the repulsive bridge correction where the atom-atom potentials contained an additional repulsive component.
In many cases one perform the bridge parameterization for different parts of the phase diagram, in particular for the density and temperature near the critical point \cite{Bondarev2008,Bondarev2010}. 
In the current paper we consider the bridge functional for the two-component systems with the fixed density and temperature of the system.
We consider only the case of the infinitely diluted solutions, because such  systems are the most often used for the prediction of the thermodynamic properties of biological molecules.
In our work we consider the simplified model, where the solvent is a liquid of Lennard-Jones balls which have the  $\sigma_{22}$  parameter coinciding with the $\sigma$ parameter if the oxygen atom of the SPC/E water model \cite{Berendsen1987} (we denote with the number 1 the sulute and with the number 2 the solvent particles).
As a solute we use the Lennard-Jones balls with diameters from $0.25\sigma_{22}$ to $2\sigma_{22}$, where $\sigma_{22}$ is the diameter of the solvent (here and below we call the $\sigma$ parameter of the LJ potential "the diameter").
The goal of the current paper is building the universal empirical bridge functional which the good coincidence with the data of the molecular dynamics (MD) simulations.
In our work we perfrom the MD simulations for all of the investigated systems and obtain the radial distribution functions (RDF).
By using the OZ equation we obtain the direct correlation functions and by using the closure relation we calculate the bridge functional for each of the systems.
We parameterize the bridge functional with the exponential function which includes two empirical parameters and determine the dependency of these parameters on the size of the solvent $\sigma_{11}$.
In such a way, we obtain the universal formula for the bridge functional for the Lennard-Jones balls of different diameter. 
At the end we compare the solutions of the Ornstein-Zernike equations with two different closures: hypper-netted-chain (HNC) closuse  which ignores the bridge functional and the closure which includes the proposed empirical bridge functional.
We show that introducing the bridge functional can essentially improve the results of the calculations and make them nearlier to the results of the MD simulations.
Out main interest is the accurate calculation of the thermodynamical and structural parameters of the aqueous solutions of bioactive compounds by using the IETL. We believe, that introducing the bridge functionals allows in many cases to substitude the comutationnaly expansive MD simulations with the cheaper IETL method, for example for the description of the interaction of osmolites with the nanoobjects and surfaces \cite{Frolov2010,Fedorov2008a,Fedorov2008,Georgi2010} and biomolecules \cite{Fedorov2009a,Fedorov2007b,Fedorov2011,Terekhova2010}.

\section{Desciption of the method}

\subsection{Molecular dynamics simulations}

\subsubsection{Description of the investigated systems}

For the molecular dynamics simulations we used the program package Gromacs 4.5.3 \cite{Hess2008}. 
We created eight systems which contain one particle with the Lennard-Jones (LJ) potential  (see the equation \EquationRef{}{eq:LJ_potential}) with eight different values of $\sigma_{11}$. The particle was solved  with infinite dilution in the LJ fluid with the fixed values $\sigma_{22}$ and $\epsilon_{22}$ (these values were the same for all systems).
We denote the solute by the digit 1 and the solvent by the digit 2. 
The solute and the solvent had the same $\epsilon$ parameter: ($\epsilon_{11}=\epsilon_{22}$).  
In our work we simulated the systems with the following values of the $\sigma_{11}$ parameter:
$\sigma_{11}= k \cdot \sigma_{22}$, где k ={0.25, 0.5, 0.75, 1.00, 1.25, 1.50, 1.75, 2.00}.
Such range of $\sigma_{11}$ parameters was used because in the popular OPLS (Optimized Potential for Liquid Simulations) force-field \cite{Jorgensen1996a} the most of $\sigma$ parameters lies in the range from 0.25 $\sigma_{22}$ to 2 $\sigma_{22}$, where $\sigma_{22}$=0.3166 nm is the LJ parameter if the water oxygen (see \FigureRef{Figure}{fig:OPLS}).
\begin{equation}\label{eq:LJ_potential} 
u^{LJ}_{ij}(r)=4\epsilon_{ij} \left( \left(  \frac{\sigma_{ij}}{r} \right)^{12}   - \left(  \frac{\sigma_{ij}}{r} \right)^{6} \right)
\end{equation} 
where $i$ and $j$ denote the type of particles.

\begin{figure}
\centering
\includegraphics[width=0.5\textwidth]{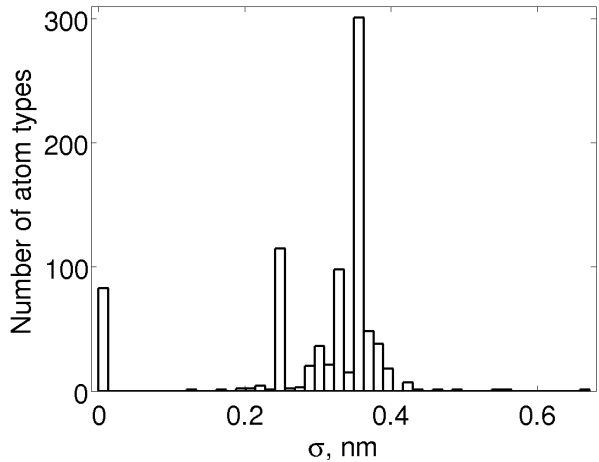}
\caption{Distibution of $\sigma$ Lennard-Jones potential parameters of the atoms from the OPLS force-filed. The data is obtained form the Gromacs 4.5.3  program package\cite{Hess2008} (file oplsaa.ff/ffnonbonded.itp).  }
\label{fig:OPLS}
\end{figure}

Mixed interaction parameters of the LJ particles of the solute and solvent were calculated using the OPLS methodology:
\begin{equation}\label{eq:opls_combining_rules} 
\begin{array}{l} {\epsilon_{12} =\sqrt{\epsilon_{11} \cdot  \epsilon_{22} } } 
\\{\sigma_{12} =\sqrt{\sigma_{11} \cdot  \sigma_{22} } } \end{array} 
\end{equation} 
Varaing the $\sigma_{11}$ we can determine the influence of the solvent at the infinite dilution in the identical solvent on the structural characteristics (solute-solvent radial distribution function (RDF)) and the shape of the bridge functional.

\subsubsection{Thermodynamical parameters, which are used in the simulations}

In the current work we are looking for the bridge functional for the follwong solvent: LJ fluid, temperature T=300K, number density 33.3294 particles/nm${}^3$ (corresponds to the water density 997.09 g/l) ,with the LJ parameters wich corresponds to the oxygen atom in the SPC/E water model ($\sigma_{22}$=0.316557 nm, $\epsilon_{22}$=0.6500194 kJ/mol) \cite{Berendsen1987}.
These parameters in the reducted units are the following: density=1.0573 particles/$\sigma_{22}^3$, T=3.8365 $\epsilon_{22}$.

\subsubsection{Parameters of the molecular dynamics symulations}

We used the "leap-frog" scheme \cite{Gromacs45} for the numerical integration of the equation of motion.
Integration step corresponded to the 0.002 ps. Cutoff radius for the LJ potential ($r_{\rm cutoff}$) was 1.2 nm, while the LJ potential was corrected to be zero at the distance $r_{\rm cutoff}$:
\begin{equation}\label{eq:shift_potential} 
\begin{array}{l l}
{ u_{ij}(r) = u^{LJ}_{ij}(r) - u^{LJ}_{ij}(r_{cutoff})} &   {, r<r_{cutoff}, } 
      \\ { u_{ij}(r) = 0} & {,r \geq r_{cutoff}}
 \end{array} 
\end{equation} 
The neighbour list was created using the "cell list" method \cite{Frenkel2002}, which is referenced in the Gromacs package as the "grid search" method \cite{Gromacs45}. The neighbour list was re-newed each tenth step with the cutoff radius 1.4 nm.
We performed the simulations in the canonical ensemble (NVT). We used the Berendsen termostat \cite{Berendsen1984} for preserving the temperature of the system (T=300K) with the parameter $\tau$=2 ps.

\subsubsection{Preparing the systems and calculationg the RDF functions}

Initially we prepared the cubic cell which contain 4168 solvent LJ particles of size 5.0007 nm$^3$ using the packmol program \cite{Martinez2009}.
Coordinates of the particles were optimized unsing the gradient descent metod \cite{Gromacs45}. 
After that the system was simulatated during the 1 ns in the NVT ensemble to establish the equilibrium.
In the final configuration one of the solvent atoms was changed to the solute atom ( LJ particle with the given $\sigma_{11}$).
The systems with different solute atoms were optimized using the gradient descent method \cite{Gromacs45}.
After that each of the systems was simulated during the 25 ns to obtain the necessary statistics.
Coordinates of the particles were saved each 0.2 ps for the further calculation of the solute-solvent RDF.
RDF was calculated using the ``g\_rdf'' tool from the Gromacs 4.5 package.

\subsection{Obtaining the bridge functional from the molecular dynamic simulation data}

\begin{figure}[h]
\centering
\includegraphics[width=0.5\textwidth]{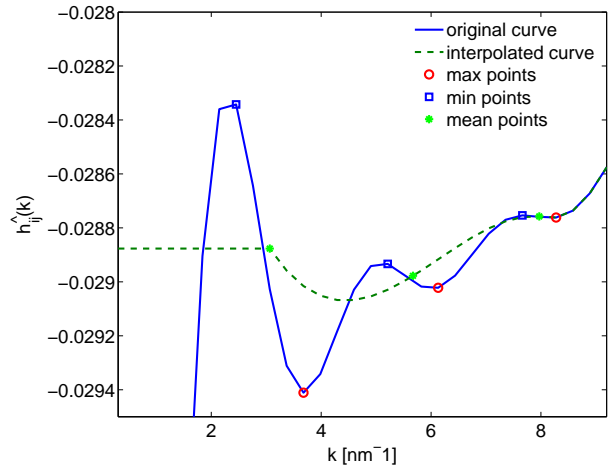}
\caption{\label{fig:denoise} 
Removing the high-frequency oscilations of the functions $\hat h_{ij}(k)$ in the vicinity of $k=0$.
In the picture is presented the function for $\sigma_{11}$=$\sigma_{22}$.
}
\end{figure}

For the bridge functional calculation we used the RDF functions wich were obtained from the MD simulations.
For the bridge functional calculation two types of RDF were used: solute-solvent RDFs ($g_{12}(r)$) and solvent-solvent RDFs ($g_{22}(r)$).
Using these functions the total correlation functions were calculated: $h_{i2}(r)=g_{i2}(r)-1, i=1,2$.
To calculate the bridge functional one also needs to know the solute-solvent direct correlation functions $c_{12}(r)$.
We used the Ornstein-Zernike equation \cite{Ornstein1914, Hansen2000} to obtain these functions:
\begin{equation}
\label{eq:c from h}
	\hat c_{12}(k) = \frac{\hat h_{12}(k)} {1+ \rho \hat h_{22}(k)}
\end{equation}
where $\rho$ is the solvent number density (number of solvent of particles in the unit volume), $\hat c_{12}(k)$, $\hat h_{12}(k)$, $\hat h_{22}(k)$ are the Fourier-Bessel transforms of the functions  $c_{12}(r)$, $h_{12}(r)$, $h_{22}(r)$ correspondingly.
The Fourier-Bessel transform of the total correlation functions $h_{ij}(r)$ is given by the following formula \cite{pyozBFT}: 
\begin{equation}
\label{eq:BFT}
   \hat h_{ij}(k) = \mathcal{F}[h_{ij}(r)] =
   \frac{4\pi}{k} \int \limits_0^{\infty} h_{ij}(r) r \sin(kr) dr
\end{equation}
The Fourier-Bessel transform on the discrete grid can produce artefacts which are seen as some oscilations of the $\hat h_{ij}(k)$ function near the $k=0$.
In the current work these artefacts were removed by the following method:
For each pair of the neighbouring local minimum and maximum points $k_{\rm min}$ and $k_{\rm max}$ of the high-frequency oscilations the value in the point $\frac{1}{2} (k_{\rm min} + k_{\rm max})$  was taken to be equal to the mean value in the extramum points:  $\hat h_{ij}( \frac{1}{2} (k_{\rm min} + k_{\rm max})) =  \frac{1}{2} (\hat h_{ij}(k_{\rm min}) + \hat h_{ij} (k_{\rm max}))$, and afterwards the function was interpolated in the vicinity of $k=0$ using the cubic splines basing on these mean values (see \FigureRef{fig:denoise}).

The  $c_{12}(r)$ can be restored from its Fourier image using the inverse Fourir-Bessel transform \cite{pyozBFT}:
\begin{equation}
\label{eq:IBFT}
  c_{12}(r) = \mathcal{F}^{-1}[\hat c_{12}(k)] = 
 \frac{1}{2\pi^2r}
 \int\limits_0^{\infty} \hat c_{12}(k) k \sin(kr) dk
\end{equation}

\begin{figure}[h]
\centering
\includegraphics[width=0.5\textwidth]{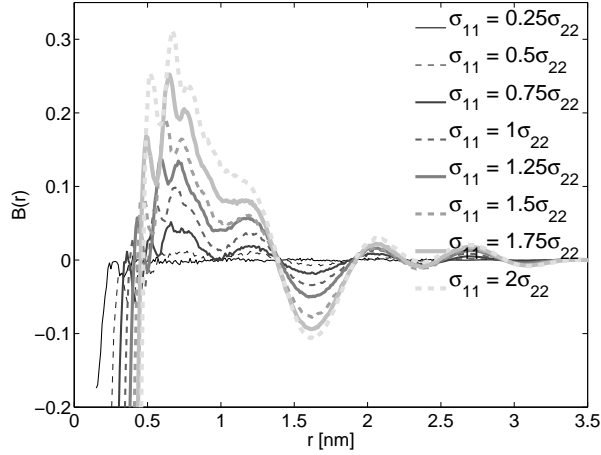}
\caption{The bridge functionals $B_{12}$ obtained from the MD simulations for the different  solute/solvent size ratios (see  \EquationRef{equation}{eq:B from MD}) }
\label{fig:bridges}
\end{figure}

The bridge functional $B_{12}(r)$ can be obtained from the closure relation using the following formula:
\begin{equation}
\label{eq:B from MD}
 B_{12}(r) = \ln g_{12}(r) + \beta U_{12}(r) - g_{12}(r) + c_{12}(r) +1
\end{equation}
Bridge functionals for the different solute/solvent size ratios are given in \FigureRef{Figure}{fig:bridges}.
Dispite the quite complicated shape of the bridge functionals in the first approximation they can be sufficiantly good approximated with the exponential function $B_{12}(a_1,a_2,r)$  with two fitting parameters $a_1$ and $a_2$:
\begin{equation}
\label{eq:B empirical}
 B_{12}(a_1,a_2,r) = -a_1 \exp (-a_2(r-r_0))
\end{equation}
The distance $r_0$ is chosen is such a way that $U_{12}(r_0)=13.8 k_BT$, where $U_{12}(r)$ is the solute-solvent LJ potential, $k_B$ is the Boltzmann constant, $T$ is the temperature.
The parameters $a_1$ and $a_2$ were chosen to minimize the difference between the RDF obtained from the MD simulations and the RDF obtained using the closure relation:
\begin{equation}
\label{eq:min}
  a_1,a_2 = {\rm argmin} || g_{12}(r) - g_{\rm closure}(a_1,a_2,r)||
\end{equation}
where $g_{\rm closure}(a_1,a_2,r) = \exp(-\beta U_{12}(r) + g_{12}(r) - c_{12}(r) + B_{12}(a_1,a_2,r) - 1  )$ and the norm is defined by the following expression:
\begin{equation}
\label{eq:norm}
  || g_{12}(r) - g_{\rm closure}(a_1,a_2,r)|| = 
  \int\limits_0^{\infty} (g_{12}(r) - g_{\rm closure}(a_1,a_2,r) )^2 dr
\end{equation}
The coefficients $a_1$ and $a_2$ were parameterized as the functions from the solute/solvent size ratio: $a_1 = a_1(\sigma_{11}/\sigma_{22})$, $a_2 = a_2(\sigma_{11}/\sigma_{22})$. 
In such a way we obtain the effective bridge functional, which depends on one parameter - the solute/solvent size ratio.
\begin{equation}
\label{eq:B(sigma)}
   B_{12}(\sigma_{11}/\sigma_{22}) = -a_1(\sigma_{11}/\sigma_{22}) \exp(-a_2(\sigma_{11}/\sigma_{22})(r-r_0))
\end{equation}

To check the effectivness of the parameterization we compared the solutions of the Ornstein-Zernike equation with the hypper netted chain closure ($B_{12}(r) =0$), the solutions of the OZ equation with the closure which includes the empirical bridge functional \EquationRef{}{eq:B(sigma)}  and the results of the MD simulations.
The Ornstein-Zernike equations were solved using the iterative algorithm.
To do this the functions $\gamma_{12}(r) = h_{12}(r) - c_{12}(r)$ were introduced.
The following algorithm was used:
the first approximation was chosen to be $\gamma_{12}^{(0)}(r)=0$.
The $(n+1)$-st approximation $\gamma_{12}^{(n+1)}(r)$ was obtained from the $n$-th approximation using the following algorithm\cite{Sergiievskyi2011}:

\begin{itemize}
\item Step 1:
$c_{12}(r)=\exp(-\beta U_{12}(r) + \gamma_{12}^{(n)}(r) + B_{12}(r) )- \gamma_{12}(r) - 1$;
\item Step 2: 
$\hat c_{12}(k) = \mathcal{F}[c_{12}(r)]$
\item Step 3:  
$\hat \gamma_{12}^{(n+1)}(k) = \rho \hat c_{12}(k) \cdot \hat h_{22}(k) $
\item Step 4: 
 $\gamma_{12}^{(n+1)}(r) = \mathcal{F}^{-1}[\hat \gamma_{12}^{(n+1)}(k)]$
\end{itemize}

\section{Results}

We performed the MD calculations for the infinitely diluted solutions of the LJ balls of dimeter $\sigma_{11}$ in the LJ fluid which consists of the balls of diameter $\sigma_{22}$. The following solute/solvent size ratios were used: $\sigma_{11}/\sigma_{22}$ = 0.25, 0.5, 0.75, 1, 1.25, 1.5, 1.75, 2.
From the MD simulation results the bridge functionals presented on the \FigureRef{Figure}{fig:bridges} were obtained by using the expressions \EquationRef{}{eq:c from h}-\EquationRef{}{eq:B from MD} .
For each ratio $\sigma_{11}/\sigma_{22}$ the obtained bridge functional was fitted with the function \EquationRef{}{eq:B empirical} where coefficients $a_1$,$a_2$  were obtained as a result of solving the minimization problem  \EquationRef{}{eq:min}. 

\begin{figure}
\centering
\includegraphics[width=0.5\textwidth]{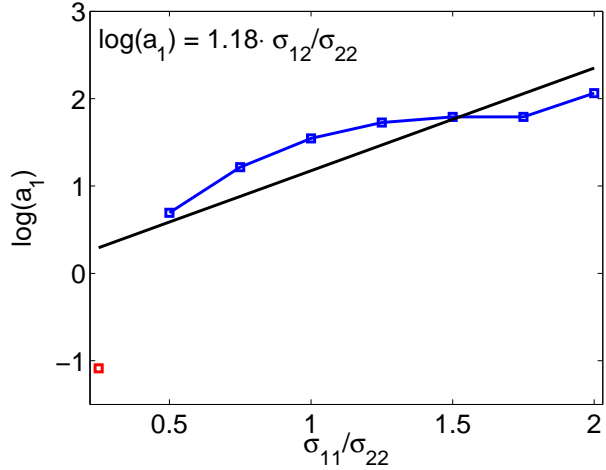}
\caption{Dependence of the $\ln(a_1)$ on the solute/solvent size ratio }
\label{fig:a1} 
\end{figure}

\begin{figure}
\centering
\includegraphics[width=0.5\textwidth]{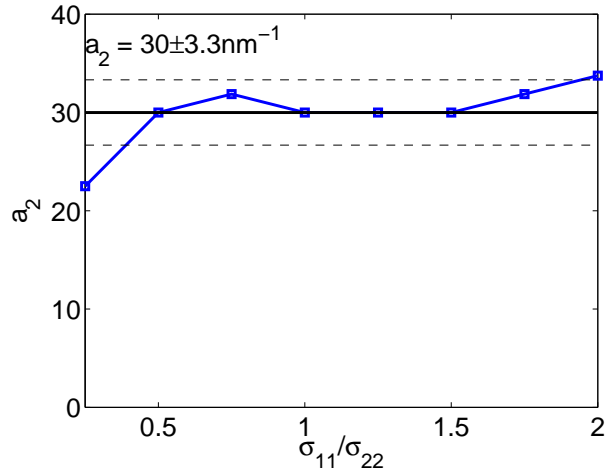}
\caption{Dependence of the $a_2$ on the solute/solvent size ratio }
\label{fig:a2} 
\end{figure}

In \FigureRef{Figure}{fig:a1} and \FigureRef{Figure}{fig:a2} one can see the dependencies of the coefficients $a_1$ and $a_2$ on the $\sigma_{11}/\sigma_{22}$ ratio.
As one can see in \FigureRef{Figure}{fig:a2}, coefficient $a_2$ weakly depend on the $\sigma_{11}/\sigma_{22}$ (standard deviation from the mean value is 3.3 nm$^{-1}$, while the mean value is 30 nm${}^{-1}$ ).
The coefficient $a_2$ can be approximated by its mean value: $a_2 \approx $ 30 nm ${}^{-1}$.
Coefficient $a_1$ have an exponential dependence from the parameter $\sigma_{11}/\sigma_{22}$ and thus  $\ln (a_1)$ linearly depend on $\sigma_{11}/\sigma_{22}$.
In \FigureRef{Figure}{fig:a1} for all $\sigma_{11}/\sigma_{22}$ ratios from 0.5 to 2 the clear linear dependence of $\ln(a_1)$ on $\sigma_{11}/\sigma_{22}$ is seen:
\begin{equation}
\label{eq:log(a1)=Csigma}
 \ln(a_1) = C\cdot \sigma_{11}/\sigma_{22}	
\end{equation}
The exception is the ratio $\sigma_{11}/\sigma_{22}$ = 0.25, which we do not use for the parameterization.
For the values pf $a_1$, which correspond to the  $\sigma_{11}/\sigma_{22}$ = 0.5, 0.75, 1, 1.25, 1.5, 1.75, 2 using the latest squares method the value of the $C$ coefficient in the equation \EquationRef{eq:log(a1)=Csigma} was determined: $C = 1.1754$.
In such a way, the final formula of the empirical bridge functional is the following:
\begin{equation}
\label{eq:B(sigma) final}	
 B_{12}(\sigma_{11}/\sigma_{22}) = - exp(-a_1(r-r_0) + C \sigma_{11} / \sigma_{22})
\end{equation}
where $a_1$=30 nm${}^{-1}$,  $C=1.1754$, $U_{12}(r_0) = 13.8 k_BT$.

\begin{figure}
\centering
\includegraphics[width=0.5\textwidth]{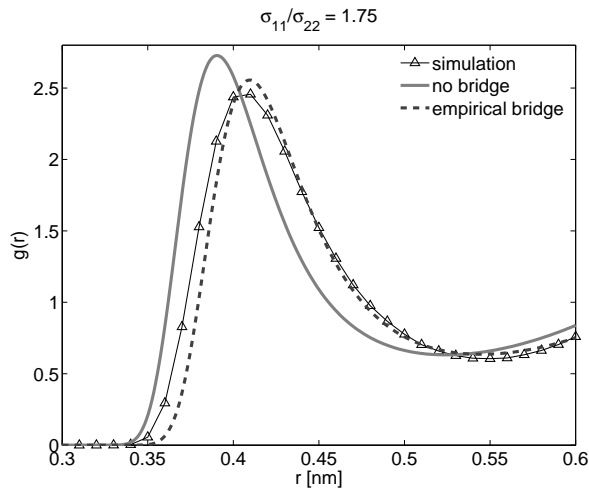}
\caption{Comparison of the RDFs obtained from the Ornstein-Zernike equation with the closure without bridge functional and with the bridge empirical functional with the MD simulation results. $\sigma_{11}/\sigma_{22}=1.75$}
\label{fig:HNCvsBR} 
\end{figure}

\begin{table}
\begin{tabular}{ | c | c |  c |  c |  c |  c | }
\hline
\multirow{2}{*}{$\sigma_{11}/\sigma_{22}$}&
\multirow{2}{*}{\parbox[c]{3cm}{peak height in MD }}& 
\multicolumn{2}{|p{4.5cm}|}{without the bridge functional} \vspace{1pt}& 
\multicolumn{2}{|p{4.5cm}|}{with the empirical bridge functional} \vspace{1pt} \\ \cline{3-6}
& & \vspace{1pt} \parbox[c]{2cm}{peak height}  & \vspace{1pt} \parbox[c]{2cm}{difference from MD} &
\vspace{1pt} \parbox[c]{2cm}{peak height}  & \vspace{1pt} \parbox[c]{2cm}{difference from MD} \\
\hline
0.25&    1.853&    1.943&    0.090&    1.831&    -0.023\\
0.5&    2.377&    2.646&    0.268&    2.473&    0.096\\
0.75&    2.490&    2.833&    0.342&    2.650&    0.160\\
1&    2.500&    2.875&    0.375&    2.701&    0.201\\
1.25&    2.521&    2.842&    0.320&    2.673&    0.152\\
1.5&    2.505&    2.793&    0.287&    2.619&    0.113\\
1.75&    2.456&    2.728&    0.272&    2.556&    0.100\\
2&    2.445&    2.662&    0.217&    2.522&    0.077\\
\hline
Average &    &    &    0.272&    &    0.109\\
\hline
\end{tabular}
\caption{\label{tab:height} 
Comparison of the heights of the first peak of RDFs which were obtained by using the different closures  with the MD simulations.
}
\end{table}

\begin{table}
\begin{tabular}{ | c |  c | c |  c |  c |  c | }
\hline
\multirow{2}{*}{$\sigma_{11}/\sigma_{22}$}&
\multirow{2}{*}{ \parbox{3cm}{peak position in MD [nm] } }& 
\multicolumn{2}{|p{4.5cm}|}{without the bridge}& 
\multicolumn{2}{|p{4.5cm}|}{with the empirical bridge }\\ \cline{3-6}
& & \vspace{2pt} \parbox{2cm}{ peak position [nm] \vspace{2pt}}  & \vspace{2pt} \parbox{2cm}{ difference from MD [nm]} &
\vspace{2pt} \parbox{2cm}{peak position [nm] } & \vspace{2pt} \parbox{2cm}{difference from MD [nm] } \\
\hline
0.25&    0.170&    0.173&    0.003&    0.179&    0.009\\
0.5&    0.230&    0.226&    -0.004&    0.230&    0.000\\
0.75&    0.280&    0.268&    -0.012&    0.274&    -0.006\\
1&    0.310&    0.304&    -0.006&    0.311&    0.001\\
1.25&    0.350&    0.336&    -0.014&    0.345&    -0.005\\
1.5&    0.380&    0.364&    -0.016&    0.378&    -0.002\\
1.75&    0.410&    0.390&    -0.020&    0.409&    -0.001\\
2&    0.430&    0.415&    -0.015&    0.437&    0.007\\
\hline
Average &    &    &    -0.011&    &    0.001\\
\hline
\end{tabular}
\caption{\label{tab:pos} Comparison of the positions of the first peak of RDFs which was obtained using the different closures with the MD simulations.}
\end{table}

To check the empirical functional \EquationRef{}{eq:B(sigma) final} for each of the ratios $\sigma_{11}/\sigma_{22}$ we solve the Ornstein-Zernike equation with two different closures: hyper-netted chain, where the bridge functional $B_{12}(r) \equiv 0$, and with the closure, which includes the empirical functional built using the formula \EquationRef{}{eq:B(sigma) final} (the $h_{22}(r)$ was taken from the MD simulation of the pure solvent).
In \FigureRef{Figure}{fig:HNCvsBR} the  RDF obtained as a result of solving the OZ equation with the closure which includes the empirical bridge functional \EquationRef{}{eq:B(sigma) final}, the RDF obtained as a result of solving the OZ equation with the closure without the bridge functional and the results of the MD simulation are shown.
The solute/solvent size ratio is $\sigma_{11}/\sigma_{22}=1.75$.
One can see, that the RDF, obtained without the bridge functional predicts incorrectly the position of the first peak of the RDF and also overestimate the height of the peak, while the RDF, obtained with the empirical bridge functional predicts correctly the position of the peak and more accurately predicts its height.
In \TableRef{Table}{tab:height}  the comparison of the first peak height obtained by using the different closures (with the bridge functional and without it) with the MD results is presented.
One can see that both of the closures overestimate the height of the first peak of RDF, but in averae the error of the closure with the bridge functional is 0.109, which is approximately 2.5 times lower than the error of the closure without the bridge.
In  \TableRef{Table}{tab:pos} the comparison of the first peak position of RDF, obtained with the two different closures (with the bridge functional and without it) with the MD results is presented.
One can see, that the closure without the bridge functional gives the systematic error of the first peak of RDF position around 0.01 nm (0.1\AA), while the closure, which includes the bridge functional, determines the first peak position with the error which is 10 times lower (0.001 nm or 0.01\AA). This error is negligable, because it is 10 times lower than the grid size which was used for the discretization of RDFs obtained from the MD simulations. 
Additionally one need to stress the importance of the accurate approximations of the first peak position of the RDF for the sovation free energy calculations, because for that application small changes in the first peak position lead to the essential errors in the calculations.

\section{Conclusions}

In the paper the new universal bridge functional for the infinitely diluted solutions of the LJ spheres of different size in the LJ fluid is proposed.
The investigated systems were the simplified model of the important case of aqueous solutions of different bioactive compounds.
$\sigma$-parameter of the solvent LJ particles was chosen to be equal to $\sigma$ parameter of the water oxygen in the SPC/E water model.
The ratios of the solute/solvent particle sizes $\sigma_{11}/\sigma_{22}$ was variaing in the range from 0.25 to 2 which approximetely corresponds to the distribution of the $\sigma$ parameters of different atoms in the OPLS force-field.

For the mentioned above systems the MD simulations were performed.
Using the RDFs, obtained from the MD simulations, and usign the Ornstein-Zernike equation, the bridge functionals were obtained, wich in turn were fitted by the exponential function \EquationRef{}{eq:B empirical}, which depends on two parameters $a_1$ and $a_2$.

Dependency of the parameters $a_1$ and $a_2$ on the  $\sigma_{11}/\sigma_{22}$ ratio was investigated.
It was shown that the $a_2$ parameter weakly depends on the size of the solute, and for all of the investigated systems it can be put to be equal $a_2=30$ nm${}^{-1}$.
It was shown that $\ln(a_1)$ correlates to the parameter $\sigma_{11}/\sigma_{22}$, and it can be estimated with a good accuracy using \EquationRef{the formula}{eq:log(a1)=Csigma}.
The final empirical bridge functional  \EquationRef{}{eq:B(sigma) final} was determined. This functional depends only on the solute/solvent particle size ratio.

Using the iterative algorithm the solutions of the Ornstein-Zernike equation for the closure with the empirical bridge functional were obtained.
It was shown that introducing the bridge functional into the closue relation reduces by 2.5 times  the error in the calculation of the first peak height of the RDF and, in contrast to the closure without the bridge, predicts accurately the position of the first peak of the RDF (with the error which is 10 times smaller than the grid discretization step). 
We have shown that introducing the empirical bridge functional into the closure relation can essentially improve the accuracy the predictions of the first peak of RDF position. 

\section{Acknowledgements}

Authors would like to acknowledge the scientific supervisor Maxim V. Fedorov for organizing the research.
Authors also would like to acknowledge the Max Planck Institute for Mathematics in the Sciences for the financial support of the research.

The molecular dynamics simulations were caried out on the suercomputer `HECToR the UK’s national high-performance computing service'', which is sipported by the following organizations: UoE HPCx Ltd  in the Univesity of Edinburg, Cray Inc. and NAG Ltd, and is financing by the ``Office of Science and Technology''  organization within the framework of ``EPSRC’s High End Computing Programme''.


\end{document}